\documentclass[referee]{raa}            

\usepackage{graphicx,times}             
\usepackage{subcaption}
\usepackage{graphicx}
\usepackage{amssymb,amsmath}

\usepackage[pagebackref=true]{hyperref}

\begin{document}

  \title{Enhancing Galaxy Classification with U-Net Variational Autoencoders for Image Denoising}

   \volnopage{Vol.0 (20xx) No.0, 000--000}      
   \setcounter{page}{1}          

   \author{S. S. Mirzoyan 
      \inst{1,2}
   }

   \institute{Center for Cosmology and Astrophysics, Alikhanyan National Laboratory, 2 Alikhanyan Brothers str., Yerevan 0036, Armenia; {\it mserg@yerphi.am}\\
        \and
             Yerevan State University, Yerevan 0025, Armenia\\
\vs\no
   {\small Received 20xx month day; accepted 20xx month day}}

\abstract{ 
AI-enhanced approaches are becoming common in astronomical data analysis, including in the galaxy morphological classification.  In this study we develop an approach that enhances galaxy classification by incorporating an image denoising pre-processing step, utilizing the U-Net Variational Autoencoder (VAE) architecture and effectively mitigating noise in galaxy images and leading to improved classification performance.
Our methodology involves training U-Net VAEs on the EFIGI dataset. To simulate realistic observational conditions, we introduce artifacts such as projected stars, satellite trails, and diffraction patterns into clean galaxy images. The denoised images generated are evaluated using Peak Signal-to-Noise Ratio (PSNR) and Structural Similarity Index (SSIM), to quantify the quality improvements.
We utilize the denoised images for galaxy classification tasks using models such as DenseNet-201, ResNet50, VGG16 and GCNN. Simulations do reveal that, the models trained on denoised images consistently outperform those trained on noisy images, thus demonstrating the efficiency of the used denoising procedure. The developed approach can be used for other astronomical datasets, via refining the VAE architecture and integrating additional pre-processing strategies, e.g. in revealing of gravitational lenses, cosmic web structures.
}

   \authorrunning{S. S. Mirzoyan} 
   \titlerunning{Enhancing Galaxy Classification with U-Net Variational Autoencoders for Image Denoising}  

   \maketitle

%
%
\section{Introduction}           
\label{sect:intro}
Galaxies within the same morphological classes are known to differ significantly by their features \cite{Mo, Sparke, Binney, Bergh}, hence the classification using more specific categories are being considered, see \cite{Doser,Luo, Gh,Tian,San}. The approach e.g. in \cite{Baillard} classifies 4458 galaxies into 18 sub-classes within 5 main classes.  Although this method offers more granularity, it is highly time and resource-consuming, requiring comprehensive examination of thousands or tens of thousands of galaxy images. The challenge becomes more pronounced when the main Hubble classes are further divided into sub-classes.


Over the past decade, machine learning (ML) has emerged as a powerful tool for astronomical data analysis \cite{Gh,Tian,San,Geo,Agui,Mas,At}. ML techniques provide superior results while saving considerable time and resources and demonstrated as being highly efficient for various types of astronomical data, often yielding better outcomes than traditional methods. For instance, \cite{Zhu} have reported a 95.2\% classification accuracy for a sample of 28,790 galaxy images from the Galaxy Zoo 2 dataset using architectures such as AlexNet, VGG, Inception, and ResNets.

Recently \cite{Genin} presented a comprehensive morphological catalog of over 340,000 galaxies from JWST NIRCam imaging across the CEERS, GOODS, and PRIMER extragalactic surveys, processed through the DAWN JWST Archive. Each source was modeled using both single Sersic and bulge+disk brightness profiles, enabling robust analysis of galaxy structure out to redshifts beyond z$>$10. Meanwhile, \cite{Kim2025} used morphological features from SDSS imaging to analyze the structural similarities among various galaxy spectral types, including strong/weak AGN, QSOs, quiescent, and star-forming galaxies, in a multidimensional feature space. Their results reveal that AGN and QSO populations are morphologically closer to quiescent galaxies, particularly in the dwarf regime, suggesting a significant role of AGN feedback in regulating star formation.

\cite{Calleja_Fuentes} created ensemble of neural networks  using the bagging ensemble method with augmented galaxy data (rotated, centered, and cropped). This work resulted 91\% accuracy when considering three galaxy types (E, S and Irr), and over 95\% accuracy for two types (E and S).
\cite{Hui} utilized a pre-trained DenseNet121 model for classification, achieving an approximately 89\% classification rate on the Galaxy10 DECals Dataset consisting of 17,736 images. Meanwhile, \cite{Pandya} employed group convolutional neural network architectures (GCNNs) equivariant to the 2D Euclidean group for galaxy morphology classification. They utilized the inherent symmetries present in galaxy images as an inductive bias, achieving a test-set accuracy of 95.52±0.18\% on the same dataset by leveraging the \( D_{16}\) symmetry of the 2D Euclidean group.

The same dataset was used for galaxy classification by \cite{qian2023performance}, where they tried various neural network architectures to address the problem. Among these, InceptionV3 achieves the highest accuracy of 83.6\%, outperforming other models across several metrics. VGG16, notable for its small convolutional kernels, balances the effectiveness of training and testing with a significant reduction in parameters, achieving 82.75\%. They report 80.78\% accuracy for ResNet50 architecture.

The ML techniques were also applied to many other classification problems, such as finding strong gravitational lenses \cite{Petrillo}, classifying galaxies in, and around, clusters, according to their projected phase-space position \cite{de2021roger}, radio galaxy classification based on continuum emission \cite{Samudre} using Siamese Networks and transfer learning techniques along with a pre-trained DenseNet model with advanced techniques like cyclical learning rate. \cite{Mirzoyan} demonstrated a new automated pipeline for searching for gravitational lenses in astronomical images, showing the method's capability to detect objects even when they are just 0.5$\sigma$ above the average background level.

In this study, we develop an innovative methodology for addressing the galaxy classification problem, focusing specifically on the application of a denoising algorithm to raw galaxy images prior to their input into the classification framework. The denoising process is conducted using convolutional variational autoencoders (VAEs), which have demonstrated significant efficacy in image denoising tasks. This additional preprocessing step was found to substantially enhance the quality of galaxy classification relative to traditional classification approaches. As distinct of the previous studies that relied on conventional methods or shallow denoising techniques, our use of U-Net VAE architecture provides an efficient scheme of noise suppression while preserving critical morphological features. Trained on realistically contaminated images from the EFIGI dataset, our model achieves superior PSNR and SSIM scores, and consistently improves the classification accuracy across multiple architectures. These results not only outperform recent benchmarks (e.g., \cite{Kim2025, Pandya, qian2023performance, Wang}) but also highlight the potential of deep generative models as a robust pre-processing tool in astronomical image analysis.

This paper is structured as follows: Section~\ref{section:dataset} provides an overview of the dataset employed in our research. In Section~\ref{section:method_processing}, we elaborate on the methodologies utilized for data processing and architectures for galaxy classification. Section~\ref{section:results} presents the experimental outcomes and offers a comprehensive discussion of their significance. Lastly, Section~\ref{section:conclusion} concludes the paper with a summary of our findings and prospective directions for future research.

\section{Dataset}
\label{section:dataset}

To analyze the efficiency of the proposed method, two different datasets have been used in our study: EFIGI (Extracted Features of Galaxies Images) and Galaxy10 DECaLS, which differ significantly in their sources, characteristics, and specific applications.

The EFIGI dataset, derived from the Third Reference Catalogue of Bright Galaxies \cite{de Vaucouleurs} and the Sloan Digital Sky Survey (SDSS), containing 4,458 galaxy images of 255x255 pixels resolution \cite{Baillard}. It provides u, g, r, i and z band CCD imaging for about 6841 ${deg}^2$ of the Northern Galactic sky complemented with spectroscopy for galaxies with r$<$17.77 magnitude over 4681 ${deg}^2$. It aims to provide a comprehensive morphological classification with 18 detailed types including Ellipticals, Spirals, and Irregulars. 

\begin{figure*}[t]
    \centering
    \includegraphics[width=\textwidth]{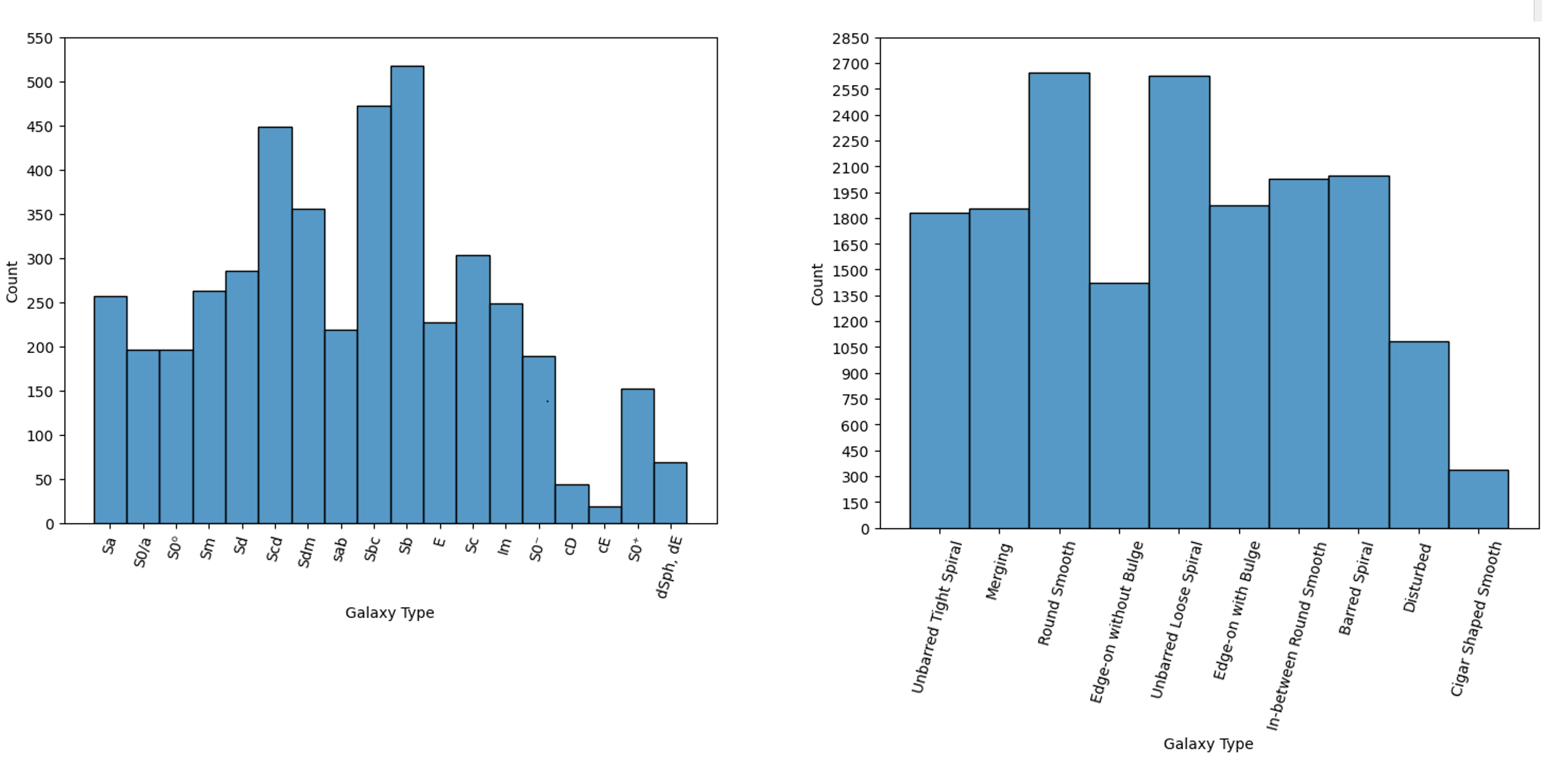}
    \caption{Galaxy distributions in EFIGI (left) and Galaxy10 DECals (rights) data sets.}
    \label{fig:gal_distrs}
\end{figure*}

As illustrated in Figure \ref{fig:gal_distrs}, the left panel, the distribution of galaxies across different morphological types demonstrates a significant imbalance within the dataset.
This dataset is enriched with expert-labeled annotations, offering fine-grained morphological features such as the presence of a bar, signs of interaction and detailed structural components.

The catalog is well sampled for the galaxies with an apparent diameter larger than ${1}$ ${arc}$ ${min}$ at the $\mu$ = 25.0 $\mathrm{mag\cdot argsec}^{-2}$ isophotal level, a total B-band magnitude brighter than about 15.5 and with less than 15000 $ km\cdot {s}^{-1}$ recession velocity \cite{Paturel2003}.

The Galaxy10 DECaLS dataset sourced from the Dark Energy Camera Legacy Survey (DECaLS), includes 17,736 colored images of galaxies in g, r, and z bands. This data set is comprised of data from the Galaxy Zoo Data Release 2 \cite{Lintott} and  DESI Legacy Imaging Surveys \cite{Lintott2011, Walmsley}. It classifies galaxies into 10 general morphological types (see Figure \ref{fig:gal_distrs} right panel for galaxy type distribution) such as completely round smooth, in-between round smooth, and edge-on galaxies. The images are provided at a resolution of 256x256 pixels, offering a less detailed but larger scale dataset ideal for deep learning applications.

\section{Methods and data pre-processing}
\label{section:method_processing}

\subsection{Image Denoising: U-net variational autoencoder} 
\label{sssec:vae_processing}

In our new approach for the galaxy classification problem, we incorporate a pre-processing step to enhance the quality of the dataset by denoising the images before training the neural network. This method leverages the power of VAEs for effective noise reduction, thus improving the subsequent classification performance.

In observational astronomy, the acquisition of galaxy images is frequently compromised by various sources of contamination. These include non-target galaxies, foreground and background stars, instrumental artifacts, and instrumental artifacts, which collectively obscure the true morphology of the target galaxies. The presence of these contaminants can introduce substantial challenges for CNNs used in astronomical image analysis. Specifically, the misclassification and misinterpretation of pixel distributions can occur when comparably large non-target objects are present within the field of view. These interfering objects can significantly alter the statistical properties of the image and mislead deep learning models, degrading their performance in tasks such as galaxy classification, morphological analysis, and distance estimation. In light of these challenges, advanced pre-processing techniques and robust training procedures are essential to mitigate the impact of contamination. Moreover, employing a VAE architecture can help in the denoising and reconstruction of cleaner galaxy images, enhancing the accuracy and reliability of subsequent analyses. By addressing these contamination sources proactively, we can improve the fidelity of image-based astronomical studies and better understand the underlying cosmic structures.

VAEs are a class of generative models that have garnered significant attention due to their ability to generate new, high-dimensional data samples by learning the underlying data distribution. Introduced by \cite{Kingma_Welling}, VAEs combine concepts from variational inference and deep learning, and are characterized by their encoder-decoder structure. Similar to ordinary autoencoders, VAEs can be employed for various applications, including image denoising, anomaly detection, and data generation. Their ability to learn meaningful latent representations makes them particularly well-suited for these tasks. 

To train the VAEs, we begin by simulating a dataset embedded with noise. This step involves taking original, high-quality galaxy images and artificially introducing noise to mimic real-world conditions that can obscure image details and impact classification accuracy. The noisy images serve as the input to the VAE, while the corresponding clean images are used as the target output.

In this study, we employ the EFIGI dataset to train a VAE for the purpose of image denoising. Subsequently, the trained VAE model will be applied to the Galaxy10 DECaLS dataset to perform denoising operations, which will facilitate further classification tasks.

The EFIGI dataset provides detailed and comprehensive descriptions of galaxies \cite{Baillard}, encompassing various morphological and structural features as well as description of environment surrounding the target galaxy. Among these descriptors, contamination and multiplicity are particularly significant for characterizing the environment of the target galaxy. The contamination refers to any additional astronomical or instrumental sources within the galaxy's field of view that impact the data quality. This includes neighboring stars and galaxies, satellite trails, and diffraction artifacts, all of which can complicate the accurate extraction of the galaxy's intrinsic properties. 

\begin{table*}[!h]
\centering
\begin{tabular}{|l|c|c|c|c|}
\hline
{\fontsize{10}{24}\selectfont \textbf{Layer Type}} & {\fontsize{10}{24}\selectfont \textbf{Filters/Units}} & {\fontsize{10}{24}\selectfont \textbf{Kernel Size}} & {\fontsize{10}{24}\selectfont \textbf{Activation}} & {\fontsize{10}{24}\selectfont \textbf{Output Shape}} \\ \hline
{\fontsize{8}{24}\selectfont  \textbf{Encoder}} \\ \hline
Input      & -   & -       & -    & (256, 256, 3)   \\ 
Conv2D     & 32  & (3, 3)  & ReLU & (256, 256, 32)  \\ 
Conv2D     & 32  & (3, 3)  & ReLU & (256, 256, 32)  \\ 
MaxPooling2D & - & (2, 2)  & -    & (128, 128, 32)  \\ 
Conv2D     & 64  & (3, 3)  & ReLU & (128, 128, 64)  \\
Conv2D     & 64  & (3, 3)  & ReLU & (128, 128, 64)  \\ 
MaxPooling2D & - & (2, 2)  & -    & (64, 64, 64)    \\ 
Conv2D     & 128 & (3, 3)  & ReLU & (64, 64, 128)   \\ 
Conv2D     & 128 & (3, 3)  & ReLU & (64, 64, 128)   \\
MaxPooling2D & - & (2, 2)  & -    & (32, 32, 128)   \\ 
Conv2D     & 256 & (3, 3)  & ReLU & (32, 32, 256)   \\ 
Conv2D     & 256 & (3, 3)  & ReLU & (32, 32, 256)   \\ 
MaxPooling2D & - & (2, 2)  & -    & (16, 16, 256)   \\
Flatten    & -   & -       & -    & 65536           \\ 
Dense (z\_mean)   & 64  & -       & -    & 64              \\ 
Dense (z\_log\_var) & 64 & -       & -    & 64              \\ 
Lambda (Sampling)  & -   & -       & -    & 64              \\ 
\hline
{\fontsize{8}{24}\selectfont  \textbf{Decoder}} \\ \hline

Input           & 64   & -  & -  & (64)  \\
Dense           & 65536 & -  & ReLU  & (65536) \\
Reshape         & -  & -  & -  & (16, 16, 256) \\
Conv2DTranspose & 256 & (3, 3), stride=2 & ReLU & (32, 32, 256) \\
Conv2DTranspose & 256 & (3, 3)  & ReLU  & (32, 32, 256) \\
Conv2DTranspose & 128 & (3, 3), stride=2 & ReLU & (64, 64, 128) \\
Conv2DTranspose & 128 & (3, 3) & ReLU & (64, 64, 128) \\
Conv2DTranspose & 64 & (3, 3), stride=2 & ReLU & (128, 128, 64)\\
Conv2DTranspose & 64 & (3, 3)  & ReLU  & (128, 128, 64)\\
Conv2DTranspose & 32 & (3, 3), stride=2 & ReLU & (256, 256, 32) \\
Conv2DTranspose & 32 & (3, 3) & ReLU & (256, 256, 32) \\
Conv2DTranspose (Output) & 3  & (3, 3)  & Sigmoid  & (256, 256, 3) \\
\hline
\end{tabular}
\caption{ U-Net Variational Autoencoder Architecture}
\label{table:U_net_vae}
\end{table*}

To ensure the integrity of our analysis, we initially filter the EFIGI dataset to select only those galaxy images exhibiting no contamination. This pre-filtering process is crucial for obtaining comparably clean galaxy images as a baseline. Consequently, our selection narrows down the dataset to 1400 galaxies from the original 4458, ensuring the exclusion of extraneous objects and artifacts. Subsequent to this filtering, we utilize the \textit{PyAutoLens} open-source package \cite{Nightingale} to simulate the presence of non-target galaxies and stars within these clean images. 

 \begin{figure*}[t]
    \centering
    \includegraphics[width=\textwidth]{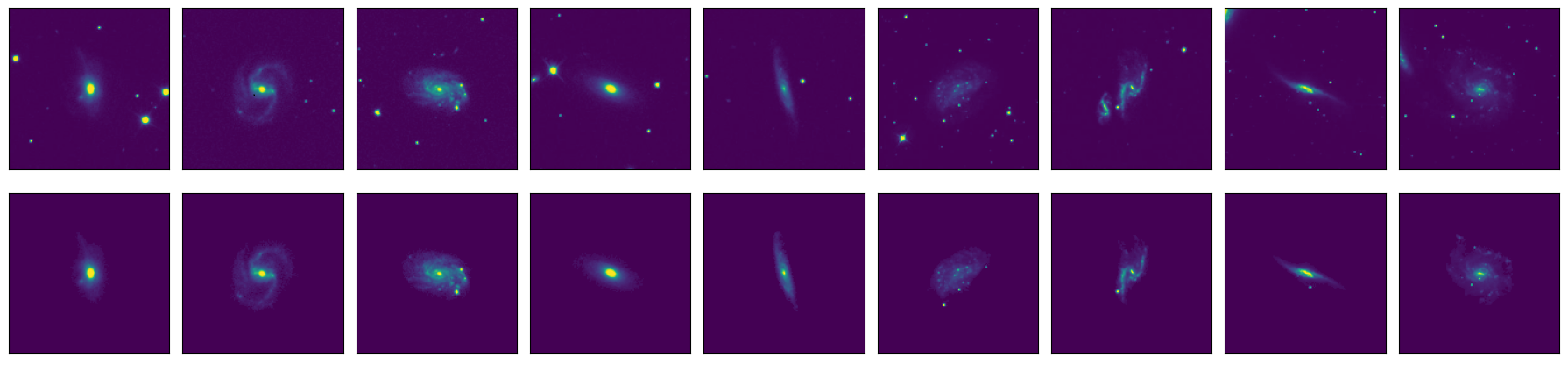}
    \caption{Denoising VAE performance on real unseen images. The first row represents real noisy images, while in the second row we show corresponding VAE denoised images.}
    \label{fig:denoising}
\end{figure*}

This simulation is designed to rigorously emulate a range of realistic astrophysical contamination. The simulations are conducted so that the new objects (non-target galaxies and stars) are randomly positioned within the image frame, ensuring comprehensive coverage of all possible spatial locations within each image. Specifically, to prevent saturation of the image frame and preserve the integrity of the simulations, the number of simulated objects in each image is kept within a random interval of 1-3. This strategic approach avoids overcrowding and ensures that the appearance of contaminants mirrors realistic observational scenarios.

The simulated galaxies and stars are generated with carefully controlled parameters to ensure astrophysical realism and varied contamination. These parameters include light profile, position, orientation, ellipticity ratio, intensity, effective radius and Sersic index.

The dataset is divided into two distinct subsets: one designated for training and on-the-fly validation, and the other reserved for testing. The training and validation subset comprises 1000 images, while the testing subset consists of 400 images.

To mitigate the risk of overfitting, our simulation efforts are confined only to the first subset. Specifically, each of the 1000 images in this subset undergoes the simulation procedure, which is repeated 10 times per image. This iterative approach results in the generation of a comprehensive augmented dataset composed of 10000 images. These repeated simulations ensure extensive coverage of the possible variations in the contaminating objects. Each simulation introduces non-target galaxies and stars with different attributes, such as varying numbers of objects (within the range of 1 to 3), diverse spatial positions, orientations, and different morphological types. By doing so, the augmented dataset encapsulates a wide spectrum of potential observational scenarios, thereby enhancing the robustness and generalizability of the trained models.

To ensure compatibility with the Galaxy10 DECals image specifications the EFIGI dataset was padded with 1 pixel along both the horizontal and vertical axes, resulting 256x256x3 pixels. Then the images were rescaled so that their pixel values fell within range of 0 to 1. This rescaling step ensures stabilizing and speeding up the training process as well as solving gradient vanishing problem.

Several experiments have demonstrated that the U-net VAE consistently exhibits superior performance as a denoising model. The U-Net Variational Autoencoder (VAE) architecture is structured into two main components: the encoder and the decoder.

The encoder is composed of four convolutional blocks, where each block contains two consecutive convolutional layers. These convolutional layers are implemented with ReLU activation functions and 'same' padding, followed by a Max Pooling layer to reduce the spatial dimensions. After the final convolutional block, a flattening layer is applied to transform the multi-dimensional output into a single-dimensional vector.

The decoder begins with a Dense layer that reshapes the latent vector back into a multi-dimensional tensor. This is followed by four blocks of transposed convolutional layers, also known as deconvolutional layers, which are used to progressively upsample the feature maps. These transposed convolutional layers reconstruct the spatial dimensions back to the original input size.

Table ~\ref{table:U_net_vae} provides a detailed representation of the architecture, including the configuration of each layer within the encoder and decoder.

The loss function utilized in this study is a combination of the reconstruction loss and the Kullback-Leibler (KL) divergence, which serves to regularize the latent space. The reconstruction loss is defined as the Binary Cross-Entropy (BCE) between the input image and the reconstructed image.

The Binary Cross-Entropy loss is given by: 
\begin{equation}
\mathcal{L}_{\text{BCE}}(x, \hat{x}) = -\sum_{i=1}^{n} \left[ x_i \log(\hat{x}_i) + (1 - x_i) \log(1 - \hat{x}_i) \right]
\end{equation}

where \textit{x} represents the true data distribution, \textit{$\hat{x}$}  denotes the predicted distribution from the decoder, and \textit{n} is the number of pixels.

The Kullback-Leibler divergence, which measures how one probability distribution diverges from a second, expected probability distribution, is defined as: 

\begin{equation}
\mathcal{L}_{\text{KL}} = -\frac{1}{2} \sum_{j=1}^{d} \left( 1 + \log(\sigma_j^2) - \mu_j^2 - \sigma_j^2 \right)
\end{equation}

where \textit{$\mu$} and \textit{$\sigma$} are the mean and standard deviation vectors of the latent variables, and d  is the dimensionality of the latent space.

The total loss function for the Variational Autoencoder can then be expressed as: 
\begin{equation}
\mathcal{L}_{\text{VAE}} = \mathcal{L}_{\text{BCE}} + \beta \mathcal{L}_{\text{KL}}
\end{equation}

 where $\beta$ is a weighting factor to balance the reconstruction loss and the KL divergence.

\begin{figure}[h]
\centering
\includegraphics[width=0.45\textwidth]{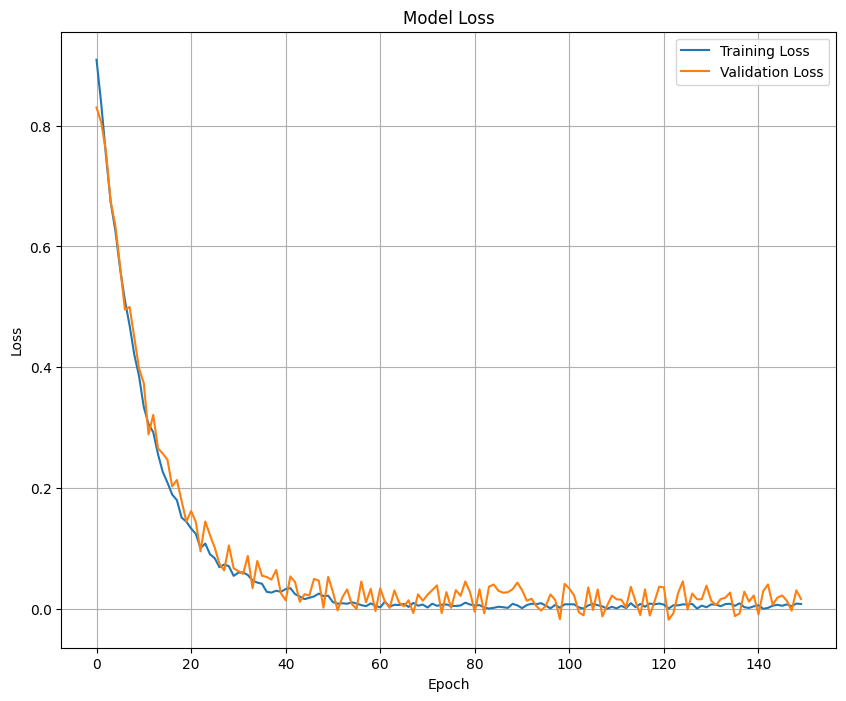}
\caption{Training and validation loss over 150 Epochs}
\label{fig:loss_plot}
\end{figure}

For the training of the U-Net VAE, we selected a batch size of 64 and conducted the training over 150 epochs. The results of the training and validation loss functions are illustrated in the Figure \ref{fig:loss_plot}.
 
The loss function results for both the training and validation datasets, evaluated after each epoch, indicate a good fit for the model's performance. Additionally, the test set loss demonstrates strong correspondence with the validation and training losses, further validating the model's generalizability and effectiveness.

In Figure \ref{fig:denoising}, the first row showcases the original noisy images, whereas the second row presents the images after the denoising process. The quality of the denoised images reflects a significant reduction in noise, highlighting the effectiveness of the U-Net VAE in enhancing image clarity.

To quantify the quality of the denoising generated by the VAE we use two metrics, Peak Signal-to-Noise Ratio (PSNR) and Structural Similarity Index (SSIM). PSNR  measures the ratio between the maximum possible power of a signal and the power of corrupting noise that affects the quality of its representation. Higher PSNR values generally indicate better quality, while SSIM measures the similarity between two images. SSIM considers changes in structural information, luminance, and contrast. Its value ranges between -1 and 1 and values closer to 1 indicate better quality.

\begin{figure*}[t]
    \centering
    \includegraphics[width=\textwidth]{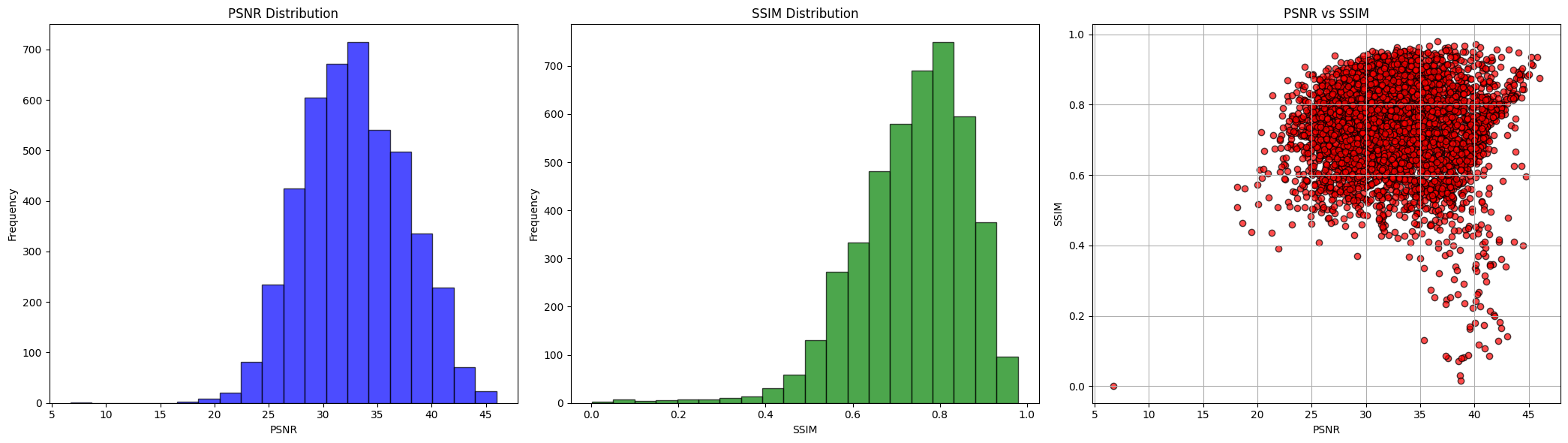}
    \caption{The figure illustrates the distribution and comparison of performance metrics for the dataset. The first plot displays the PSNR distribution, highlighting the range and variability of PSNR values within the dataset. The second plot shows the SSIM distribution, representing the spread and consistency of SSIM values. The third plot visualizes the relationship between PSNR and SSIM by plotting PSNR against SSIM, providing insights into the correlation between the two metrics.}
    \label{fig:psnr_ssim}
\end{figure*}

The quality of the denoising is depicted in  Figure \ref{fig:psnr_ssim}. It consists of three plots, each providing significant insights into the quality metrics of the image denoising. The first plot displays the distribution of PSNR values. The PSNR values exhibit an average of 32.9 dB with a standard deviation of 4.6 dB. This suggests a generally high fidelity of the denoised images relative to the original images, with variations in the fidelity across different images.

The second plot illustrates the distribution of SSIM values. The SSIM values present an average of 0.78 and a standard deviation of 0.12. These statistics imply that most denoised images maintain a considerable degree of structural similarity to the original images, although there is some variation in the structural integrity preserved.
The third plot is a scatter plot of PSNR versus SSIM values. A prominent concentration of data points in the upper right quadrant indicates that a substantial number of images exhibit both high fidelity (as evidenced by high PSNR values) and high structural similarity (as indicated by high SSIM values). This suggests that the denoising technique effectively balances both noise reduction and structural preservation.

\subsection{Galaxy classification}

With the denoised dataset prepared, we now focus on the architecture and methodology used for the galaxy classification task. The objective of this task is to accurately categorize galaxies from Galaxy10 DECals dataset into their respective morphological classes.

Our objective is to demonstrate the efficacy of our proposed approach, which involves the denoising of galaxy images using a U-net VAE prior to the galaxy classification task. To validate the effectiveness of this method, we conducted several experiments utilizing various classification models, including DenseNet-201, ResNet50 and VGG16.

To set a benchmark for our classification performance, we adopted the methodology proposed by \cite{Pandya} in their work  as well. This study employed E(2) equivariant neural networks, which demonstrated state-of-the-art accuracy in galaxy morphology classification by effectively leveraging symmetries inherent in galaxy images. 

In this research, we employ Group Convolutional Neural Networks (GCNNs)  \cite{Cohen}. GCNNs leverage the symmetries inherent in the data as an inductive bias, facilitating increased weight sharing relative to conventional CNNs. This approach enhances robustness to geometric perturbations such as rotations and  reflections, as highlighted in prior studies \cite{Ciprijanovi}, \cite{Ciprijanovi22}, \cite{Dumont}.

Similar to \cite{Pandya} we utilized the software package \textit{escnn} \cite{Cesa, Weiler} to construct GCNNs utilizing the \( D_{16}\) symmetry group of the 2D Euclidean plane. The \( D_{16}\) symmetry, also known as the dihedral group of order 16, encapsulates the symmetries of a regular polygon with 8 sides. This group consists of 16 elements, including 8 rotations and 8 reflections. By incorporating \( D_{16}\) symmetry, our GCNNs are designed to be equivariant to a comprehensive set of geometric transformations, making them particularly robust to both rotations and reflections.

\begin{figure}[h]
\centering
\includegraphics[width=0.45\textwidth]{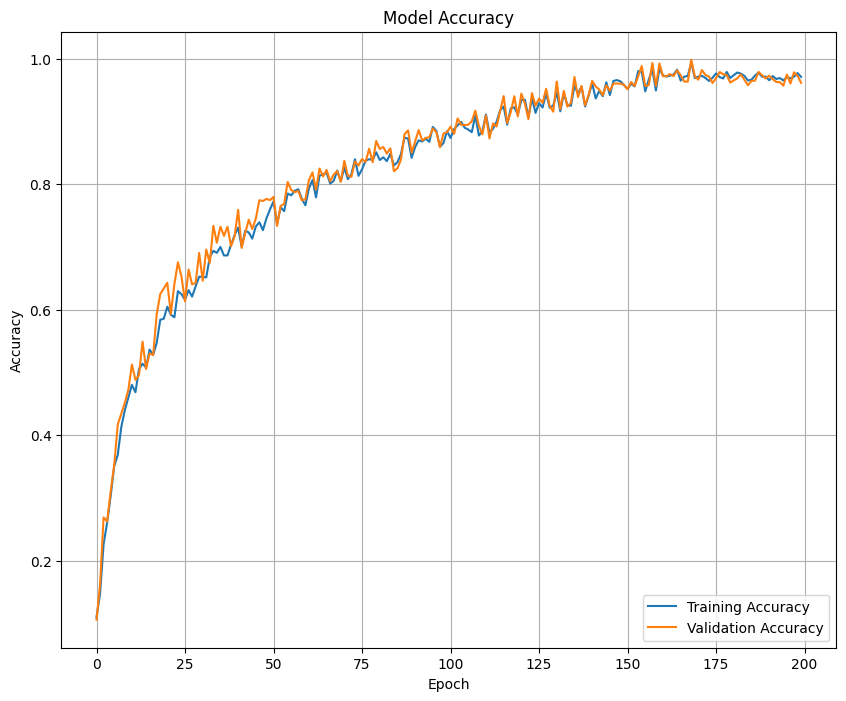}
\caption{Training and validation accuracy over 200 Epochs}
\label{fig:accuracy}
\end{figure}

The Galaxy10 DECals dataset was partitioned into three subsets: 12,736 images for training, 2,500 images for validation, and 2,500 images for testing. All images are normalized according to the procedure outlined in ~\autoref{sssec:vae_processing}. To identify the most optimal configuration for our classification model, we conducted a series of experiments testing various combinations of optimization algorithms, batch sizes, loss functions, and learning rates.
Through rigorous experimentation, we determined that the following configuration yielded the most promising results: 128 batch size, 0.0001 for learning rate, categorical cross entropy  for the loss and Adagrad for an optimizer. Figure \ref{fig:accuracy} illustrates the evolution of both training and validation accuracies over the course of 200 epochs. One can easily observe that the training and validation accuracies are very consistent with each other and converge smoothly, indicating the robustness of our model and the efficacy of our approach.

\begin{table*}[ht]
    \centering
    \resizebox{\textwidth}{!}{%
    \begin{tabular}{|l|c|c|c|c|c|c|c|c|c|c|}
        \hline
        & \textbf{Dist.} & \textbf{Mer.} & \textbf{Rnd. Sm.} & \textbf{Rnd. Sm. In-B.} & \textbf{Cig. Sh. Sm.} & \textbf{Bar. Sp.} & \textbf{Unb. T. Sp.} & \textbf{Unb. L. Sp.} & \textbf{Edg. w/o B.} & \textbf{Edg. w/ B.} \\
        \hline
        \textbf{Disturbed} & 148 & 1 & 0 & 0 & 1 & 1 & 0 & 1 & 0 & 1 \\
        \hline
        \textbf{Merging} & 0 & 240 & 1 & 0 & 0 & 1 & 1 & 0 & 0 & 1 \\
        \hline
        \textbf{Round Smooth} & 1 & 0 & 359 & 2 & 0 & 2 & 3 & 1 & 1 & 0 \\
        \hline
        \textbf{In-between Round Smooth} & 0 & 1 & 1 & 292 & 0 & 1 & 0 & 2 & 0 & 1 \\
        \hline
        \textbf{Cigar Shaped Smooth} & 1 & 0 & 1 & 0 & 45 & 0 & 1 & 0 & 0 & 0 \\
        \hline
        \textbf{Barred Spiral} & 0 & 0 & 0 & 2 & 0 & 267 & 1 & 2 & 0 & 3 \\
        \hline
        \textbf{Unbarred Tight Spiral} & 1 & 0 & 1 & 1 & 2 & 2 & 265 & 0 & 1 & 0 \\
        \hline
        \textbf{Unbarred Loose Spiral} & 0 & 1 & 1 & 1 & 0 & 0 & 0 & 369 & 0 & 1 \\
        \hline
        \textbf{Edge-on without Bulge} & 1 & 0 & 1 & 1 & 1 & 0 & 0 & 1 & 188 & 1 \\
        \hline
        \textbf{Edge-on with Bulge} & 0 & 0 & 0 & 1 & 1 & 1 & 1 & 0 & 1 & 268 \\
        \hline
    \end{tabular}%
    }
    \caption{Confusion Matrix with galaxy types}
    \label{table:confusion_matrix}
\end{table*}

\section{Results of analysis}
\label{section:results}

To enhance the quality of galaxy images prior to classification, we employed a U-Net VAE for image denoising. This approach leverages the architectural advantages of U-Net combined with the generative capabilities of VAEs to effectively reduce noise while preserving essential image features.
The PSNR and SSIM values were calculated to quantify the denoising performance: the PSNR demonstrated an average of 32 dB with a standard deviation of 4.6 dB, while the SSIM achieved a mean of 0.78 and an average of 0.12. Figure \ref{fig:psnr_ssim} illustrates the effectiveness of our denoising approach.
Visual inspections further corroborate the quantitative findings. Figure \ref{fig:denoising} illustrates examples of noisy images (upper row) along with their denoised counterparts (lower row).

Using the denoised images, we proceeded to train and evaluate our galaxy classification model. The classifier's performance was assessed using standard metrics: accuracy, precision, recall, and F1-score, which were calculated by confusion matrix (see Table ~\ref{table:confusion_matrix}).

For comparative analysis, we conducted image classification using several different  architectures on the denoised dataset. Specifically, we employed DenseNet-201, ResNet50, VGG16, and GCNNs. This comparative study aims to evaluate the performance of these architectures in classifying galaxy images post-denoising.

Table ~\ref{table:classification_results} indicates the performance metrics of our proposed denoising and classification workflow different approaches. GCNN outperforms all other models in terms of accuracy, precision, recall, and F1-score.

These results underscore the importance of  classification model for improving the accuracy of galaxy classification tasks, alongside effective image pre-processing techniques. 
It is important to stress that all four classification experiments performed with the denoised dataset outperformed their counterparts' results from \cite{qian2023performance}, \cite{Wang} and \cite{Pandya}. This highlights the effectiveness of our denoising approach in enhancing the classification performance.

\begin{table}[h!]
\centering
\begin{tabular}{lcccc}
\textbf{Model} & \textbf{Accuracy} & \textbf{Precision} & \textbf{Recall} & \textbf{F1-Score} \\
\textbf{GCNN} & 97.45 & 97.05 & 96.22 & 96.4 \\
\textbf{VGG} & 89.57 & 86.0 & 87.44 & 86.1 \\
\textbf{ResNet} & 87.2 & 87.55 & 82.5 & 83.35 \\
\textbf{DenseNet}& 91.2 & 90.07 & 89.71 & 90.02 \\

\end{tabular}
\caption{Performance metrics of different models on the galaxy classification dataset. The GCNN model demonstrates the highest accuracy and F1-Score, indicating its superiority in classifying galaxy images compared to VGG, ResNet, and DenseNet.}
\label{table:classification_results}
\end{table}

\section{Summary and conclusions}
\label{section:conclusion}
In this study, we developed a novel approach for the galaxy classification problem that incorporates a pre-processing step to enhance dataset quality through image denoising by leveraging the power of U-net VAEs, and efficiently reducing the noise in galaxy images, thereby improving the performance of subsequent classification tasks.

Our procedure began by training the VAEs using the EFIGI dataset, which offers comprehensive descriptions of galaxies, including crucial morphological and structural features. We simulated realistic noise by introducing artifacts such as neighboring stars and galaxies, satellite trails, and diffraction patterns into originally uncontaminated galaxy images. This approach ensured a robust training set for the VAEs by mimicking real-world observational conditions. The denoised images generated by the VAEs exhibited superior quality, as quantified by metrics such as Peak Signal-to-Noise Ratio (PSNR) and Structural Similarity Index (SSIM). 

Using these denoised images, we proceeded with galaxy classification tasks employing several models, including DenseNet-201, ResNet50, VGG16 and GCNN. The results demonstrated that models trained on denoised images consistently outperformed those trained on noisy images, showcasing the efficacy of our denoising approach. Notably, all classification experiments with the denoised dataset surpassed the results reported in previous studies~\cite{qian2023performance},~\cite{Wang}, and~\cite{Pandya}.

The U-Net VAE architecture proved particularly efficient in this context. Visual inspections and quantitative metrics confirmed that the model significantly reduced noise while preserving essential image features, thus enhancing the clarity and reliability of the galaxy images. The performance improvements noted in classification tasks emphasize the importance of effective pre-processing steps in astronomical image analysis.

In conclusion, our integrated approach of using U-Net VAEs for image denoising prior to galaxy classification has revealed its efficiency in both image quality and classification accuracy. This denoising technique can be applied to other astronomical datasets, refining the VAE architecture, and combining it with additional pre-processing methods to further enhance the revealing not only the galaxy morphological features but also of other cosmic structures.

\begin{acknowledgements}
{We thank the anonymous referee for insightful comments. We acknowledge the use of the EFIGI and Galaxy10 DECaLS datasets, which were instrumental in the development and validation of this research.}
\end{acknowledgements}

\label{lastpage}

\end{document}